\documentclass[twocolumn,aps,prl,showpacs]{revtex4}
\usepackage[T1]{fontenc}
\usepackage[latin9]{inputenc}
\usepackage{amsmath}
\usepackage{amssymb}
\usepackage{graphicx}
\usepackage{esint}

\makeatletter
\@ifundefined{textcolor}{}
{%
 \definecolor{BLACK}{gray}{0}
 \definecolor{WHITE}{gray}{1}
 \definecolor{RED}{rgb}{1,0,0}
 \definecolor{GREEN}{rgb}{0,1,0}
 \definecolor{BLUE}{rgb}{0,0,1}
 \definecolor{CYAN}{cmyk}{1,0,0,0}
 \definecolor{MAGENTA}{cmyk}{0,1,0,0}
 \definecolor{YELLOW}{cmyk}{0,0,1,0}
}

\makeatother

\begin{document}

\title{Gapless topological Fulde-Ferrell superfluidity in spin-orbit coupled
Fermi gases}

\author{Ye Cao$^{1,2}$}

\author{Shu-Hao Zou$^{3}$}

\author{Xia-Ji Liu$^{1}$}

\author{Su Yi$^{3}$}

\author{Gui-Lu Long$^{2,4,5}$}

\author{Hui Hu$^{1}$}

\email{hhu@swin.edu.au}

\affiliation{$^{1}$Centre for Quantum and Optical Science, Swinburne University
of Technology, Melbourne 3122, Australia}

\affiliation{$^{2}$State Key Laboratory of Low-dimensional Quantum Physics and
Department of Physics, Tsinghua University, Beijing 100084, P. R.
China}

\affiliation{$^{3}$State Key Laboratory of Theoretical Physics, Institute of
Theoretical Physics, Chinese Academy of Sciences, Beijing 100190,
P. R. China}

\affiliation{$^{4}$Collaborative Innovation Center of Quantum Matter, Beijing
100084, P. R. China}

\affiliation{$^{5}$Tsinghua National Laboratory for Information Science and Technology,
Beijing 100084, P. R. China}
\begin{abstract}
Topological superfluids usually refer to a superfluid state which
is gapped in the bulk but metallic at the boundary. Here we report
that a gapless, topologically non-trivial superfluid with inhomogeneous
Fulde-Ferrell pairing order parameter can emerge in a two-dimensional
spin-orbit coupled Fermi gas, in the presence of both in-plane and
out-of-plane Zeeman fields. The Fulde-Ferrell pairing - induced by
the spin-orbit coupling and in-plane Zeeman field - is responsible
for this gapless feature. This exotic superfluid has a significant
Berezinskii-Kosterlitz-Thouless (BKT) transition temperature and has
robust Majorana edge modes against disorder owing to its topological
nature.
\end{abstract}


\maketitle
Over the past few years, exotic pairing mechanism has gained widespread
concern, making it spring up in a wide range of areas from astrophysics,
solid-state physics to ultracold atomic physics, to name a few \cite{Casalbuoni2004,Uji2006,Kenzelmann2008,Liao2010,Gerber2014}.
The spatially modulated Fulde-Ferrell-Larkin-Ovchinnikov (FFLO) state
plays a key role in this mechanism \cite{Fulde1964,Larkin1964} and
could emerge in spin-imbalanced systems, in which the Bardeen-Cooper-Schrieffer
(BCS) state may become unstable against the pairing with finite center-of-mass
momentum. Taking the advantage of high controllability \cite{Zwierlein2006,Partridge2006},
ultracold atomic Fermi gases are ideal table-top systems for pursuing
the FFLO state \cite{Sheehy2006,Hu2006,Gubbels2013}. Indeed, strong
evidence for FFLO superfluidity has been seen in a Fermi cloud of
$^{6}$Li atoms confined in one dimensional harmonic traps \cite{Orso2007,Hu2007,Liao2010}.
Most recently, motived by the realization of synthetic spin-orbit
coupling (SOC) in cold atoms \cite{Lin2011,Wang2012,Cheuk2012}, FF
superfluidity is also argued to be observable in spin-orbit coupled
atomic Fermi gases \cite{Zheng2013,Wu2013,Liu2013a,note1}. It is
induced by the combined effect of SOC and in-plane Zeeman field, which
leads to the deformation of the Fermi surfaces \cite{Barzykin2002,Dong2013,Shenoy2013}.

Topological insulators and superconductors have been another hot research
area in recent years \cite{Hasan2010,Qi2011}. These materials are
gapped in the bulk but metallic at the boundary, supporting the prerequisites
of some crucial physical realities, for example, the non-Abelian anyons
used to form quantum gates in fault-tolerant quantum computation \cite{Nayak2008}.
It is now widely believed that topological superconductors (or superfluids)
could acquire zero-energy edge states known as Majorana fermions -
non-Abelian particles that are their own antiparticles - which are
still mysterious and not observed distinctly in recent experiments
\cite{Mourik2012,Williams2012,Rokhinson2012}. First proposed to be
realizable in chiral \textit{p}-wave superconductors, Majorana fermions
could also exist in \textit{s}-wave superconductors with SOC \cite{Zhang2008,Sato2009,Sau2010,Oreg2010}.
In the context of ultracold atomic Fermi gases, there are already
some theoretical works detailing the emergence and stability of topological
superfluids in the presence of an out-of-plane Zeeman field \cite{Liu2012a,Liu2012b,Wei2012,Ruhman2014}.
Interestingly, topological order is compatible with inhomogeneous
FF superfluidity \cite{Chen2013,Liu2013b,Qu2013,Zhang2013}. In the
case of a two-dimensional (2D) atomic Fermi gas with Rashba SOC and
both in-plane and out-of-plane Zeeman fields, a gapped topological
FF superfluid has been predicted very recently \cite{Qu2013,Zhang2013}.

\begin{figure}[t]
\begin{centering}
\includegraphics[clip,width=0.44\textwidth]{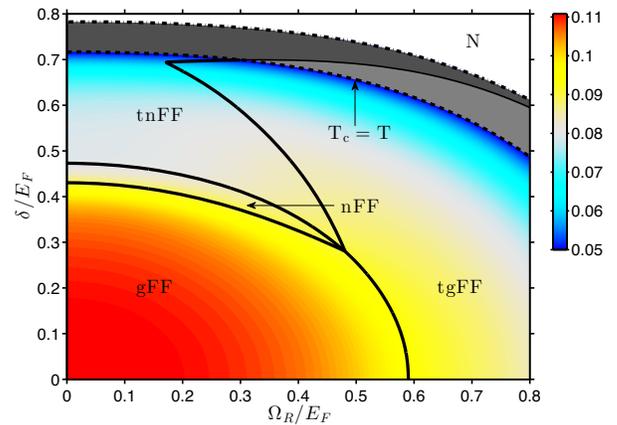} 
\par\end{centering}

\caption{(Color online) Low-temperature phase diagram of a Rashba spin-orbit
coupled 2D Fermi gas at $E_{b}=0.2E_{F}$, $\lambda=E_{F}/k_{F}$
and $T=0.05T_{F}$. By tuning the out-of-plane and in-plane Zeeman
fields, $\Omega_{R}$ and $\delta$, the system may evolve from a
gapped FF (gFF) to a nodal FF (nFF), then a gapless topological FF
(tnFF), and finally to a gapped topological FF (tgFF). Above the dot-dashed
line, the order parameter $\Delta$ becomes less than $0.001E_{F}$
(therefore indicated as a normal phase N). The dashed line indicates
the boundary where the critical BKT temperature is reached. In between,
the pseudogap regime is highlighted by the shadowed area. The color
map shows the phase stiffness $\pi\mathcal{J}/2$, in units of $E_{F}$.}

\label{fig1} 
\end{figure}

In this Letter, we report the emergence of \emph{gapless} topological
FF superfluid in the same 2D setting, as indicated by the ``tnFF''
phase in Fig. \ref{fig1}, which intervenes between the topologically
trivial (gFF or nFF) and the gapped topologically non-trivial FF states
(tgFF), and occupies a sizable parameter space. Furthermore, its existence
is not restricted to the pure Rashba or Dresselhaus SOC. Any type
of SOC with an unequal weight in the Rashba and Dresselhaus components
can support such an exotic superfluid. These findings are remarkable,
as commonly topological superfluids are believed to have an energy
gap in the bulk \cite{Hasan2010,Qi2011}. Here our goal is to understand
why the gapless energy structure coexists with the topological order,
over a broad range of parameters, at both zero and finite temperatures.
Understanding this may shed lights on designing new gapless topological
materials in solid-state systems \cite{Alicea2010}. 

We start by considering the model Hamiltonian of a 2D spin-orbit coupled
two-component Fermi gas with the SOC $\lambda_{x}\hat{k}_{x}\sigma_{x}+\lambda_{y}\hat{k}_{y}\sigma_{y}$,
the in-plane ($\delta$) and out-of-plane ($\Omega_{R}$) Zeeman fields,
$\text{\ensuremath{\mathcal{H}}}=\int d{\bf r}[\mathcal{H}_{0}+\mathcal{H}_{int}]$,
where $\mathcal{H}_{0}$ is the single-particle Hamiltonian,

\begin{equation}
{\cal H}_{0}=\left[\psi_{\uparrow}^{\dagger}\left(\mathbf{r}\right),\psi_{\downarrow}^{\dagger}\left(\mathbf{r}\right)\right]\left[\begin{array}{cc}
\xi_{{\bf k}+} & \Lambda_{{\bf k+}}^{\dagger}\\
\Lambda_{{\bf k}+} & \xi_{{\bf k-}}
\end{array}\right]\left[\begin{array}{c}
\psi_{\uparrow}\left(\mathbf{r}\right)\\
\psi_{\downarrow}\left(\mathbf{r}\right)
\end{array}\right],\label{eq:spHami}
\end{equation}
and $\text{\ensuremath{\mathcal{H}}}_{int}=U_{0}\psi_{\uparrow}^{\dagger}({\bf r})\psi_{\downarrow}^{\dagger}({\bf r})\psi_{\downarrow}({\bf r})\psi_{\uparrow}({\bf r})$
is the interaction Hamiltonian with a pairwise contact interaction
of strength $U_{0}$. Here we have used the notations, $\xi_{{\bf k\pm}}\equiv\hat{\xi}_{{\bf k}}\pm\Omega_{R}\equiv-\hbar^{2}\nabla^{2}/(2m)-\mu\pm\Omega_{R}$
with the atomic mass $m$ and chemical potential $\mu$, and $\Lambda_{{\bf k\pm}}\equiv\lambda_{x}\hat{k}_{x}+i\lambda_{y}\hat{k}_{y}\pm\delta$,
where $\hat{k}_{x}=-i\hbar\partial_{x}$ and $\hat{k}_{y}=-i\hbar\partial_{y}$
are momentum operators. $\psi_{\sigma}^{\dagger}({\bf r})$ ($\psi_{\sigma}({\bf r})$)
is the field operator for creating (annihilating) an atom with pseudo-spin
state $\sigma\in(\uparrow,\downarrow)$ at ${\bf r}$. We have explicitly
adopted a general form of SOC including both Rashba (i.e., $\lambda_{x}=\lambda_{y}$)$ $
and Dresselhaus ($\lambda_{x}=-\lambda_{y}$) SOCs. In the recent
experiments \cite{Wang2012,Cheuk2012}, only the SOC with equally
weighted Rashba and Dresselhaus components has been realized. The
in-plane ($\delta$) and out-of-plane ($\Omega_{R}$) Zeeman fields
can be created depending on the detailed experimental realization.
In the interaction Hamiltonian, the bare interaction strength $U_{0}$
is to be regularized as $U_{0}^{-1}=-\mathcal{S}^{-1}\sum_{\mathbf{k}}1/(E_{b}+\hbar^{2}\mathbf{k}^{2}/m)$,
where $\mathcal{S}$ is the area of the system and $E_{b}$ is the
two-particle binding energy.

In the presence of an in-plane Zeeman term $\delta\sigma_{x}$ in
the Hamiltonian, it is known that a finite momentum pairing will arise
along the \textit{x}-direction \cite{Liu2013a}. Focusing on a FF-like
order parameter $\Delta({\bf r})=-U_{0}\left\langle \psi_{\downarrow}(\mathbf{{\bf r}})\psi_{\uparrow}(\mathbf{{\bf r}})\right\rangle =\Delta e^{iQx}$
at the mean-field level, the interaction Hamiltonian can be approximated
by ${\cal H}_{int}\simeq-[\Delta(\mathbf{{\bf r}})\psi_{\uparrow}^{\dagger}(\mathbf{{\bf r}})\psi_{\downarrow}^{\dagger}(\mathbf{{\bf r}})+\textrm{H.c.}]-\left|\Delta({\bf r})\right|^{2}/U_{0}$.
By introducing the Nambu spinor $\Phi(\mathbf{{\bf r}})\equiv[\psi_{\uparrow},\psi_{\downarrow},\psi_{\uparrow}^{\dagger},\psi_{\downarrow}^{\dagger}]$$^{T}$,
the total Hamiltonian can be rewritten in a compact form, $\mathcal{H}=(1/2)\int d{\bf {\bf r}}\Phi^{\dagger}(\mathbf{{\bf r}})\mathcal{H}_{BdG}\Phi(\mathbf{{\bf r}})-\mathcal{S}\Delta^{2}/U_{0}+\sum_{\mathbf{k}}\hat{\xi}_{\mathbf{k}}$,
where $\mathcal{H}_{BdG}$ is given by,

\begin{equation}
\mathcal{H}_{BdG}\equiv\left[\begin{array}{cccc}
\xi_{{\bf k+}} & \Lambda_{{\bf k}+}^{\dagger} & 0 & -\Delta\left(\mathbf{{\bf r}}\right)\\
\Lambda_{{\bf k+}} & \xi_{{\bf k-}} & \Delta\left(\mathbf{{\bf r}}\right) & 0\\
0 & \Delta^{*}\left(\mathbf{{\bf r}}\right) & -\xi_{{\bf k+}} & \Lambda_{{\bf k-}}\\
-\Delta^{*}\left(\mathbf{{\bf r}}\right) & 0 & \Lambda_{{\bf k-}}^{\dagger} & -\xi_{{\bf k-}}
\end{array}\right].\label{eq:BdGHami}
\end{equation}
It is straightforward to diagonalize the above Bogoliubov Hamiltonian
$\mathcal{H}_{BdG}\Phi_{\mathbf{k\eta}}^{\nu}(\mathbf{{\bf r}})=E_{\mathbf{k\eta}}^{\nu}\Phi_{\mathbf{k\eta}}^{\nu}(\mathbf{{\bf r}})$
with quasiparticle energy $E_{\mathbf{k}\eta}^{\nu}$ and quasiparticle
wave-function $\Phi_{\mathbf{k\eta}}^{\text{\ensuremath{\nu}}}(\mathbf{{\bf r}})=1/\sqrt{\mathcal{S}}e^{i\mathbf{k\cdot}\mathbf{{\bf r}}}[u_{\mathbf{\eta\uparrow}}^{\nu}e^{+iQx/2},u_{\mathbf{\eta\downarrow}}^{\nu}e^{+iQx/2},v_{\mathbf{\eta\uparrow}}^{\nu}e^{-iQx/2},v_{\mathbf{\eta\downarrow}}^{\nu}e^{-iQx/2}]^{T}$,
where $\nu\in(+,-)$ represents the particle ($+$) or hole ($-$)
branch, and $\text{\ensuremath{\eta\in}(1,2)}$ denotes the upper
($1$) or lower ($2$) helicity branch \cite{Hu2011}. The mean-field
thermodynamic potential $\Omega$ at the temperature $T$ is then
given by,
\begin{eqnarray}
\Omega_{\textrm{mf}} & \textrm{=} & \frac{1}{2}\sum_{\mathbf{k}}\left(\xi_{\mathbf{k}+\mathbf{Q}/2}+\xi_{\mathbf{k}-\mathbf{Q}/2}\right)-\frac{1}{2}\sum_{\mathbf{k\eta}}|E_{\mathbf{k}\eta}^{\nu=+}|\nonumber \\
 &  & -k_{B}T\sum_{\mathbf{k\eta}}\ln\left(1+e^{-|E_{\mathbf{k}\eta}^{\nu=+}|/k_{B}T}\right)-\mathcal{S}\frac{\Delta^{2}}{U_{0}},\label{eq:Omega}
\end{eqnarray}
where $\xi_{\mathbf{k}\pm\mathbf{Q}/2}=\hbar^{2}(\mathbf{k}\pm\mathbf{Q}/2)^{2}/(2m)-\mu$.
Taking the advantage of inherent particle-hole symmetry in the Nambu
spinor representation \cite{Liu2013a}, only the eigenvalues in the
particle branch ($\nu=+$) is necessary in the calculation of the
thermodynamic potential. For a given set of parameters (i.e., the
temperature $T$, binding energy $E_{b}$ etc.), different mean-field
phases can be determined using the self-consistent stationary conditions:
$\partial\Omega_{\textrm{mf}}/\partial\Delta=0$, $\partial\Omega_{\textrm{mf}}/\partial Q=0$,
as well as the conservation of total atom number, $N=-\partial\Omega_{\textrm{mf}}/\partial\mu$.
At a given temperature, the ground state has the lowest free energy
$F=\Omega_{\textrm{mf}}+\mu N$. To characterize its stability, we
calculate the superfluid density tensor by applying a phase twist
to the order parameter $\Delta({\bf r})=\Delta e^{iQx+i\mathbf{q}\cdot\mathbf{r}}$,
which boosts the system with a superfluid flow with a velocity $\mathbf{v}_{s}=\hbar\mathbf{q}/2m$.
The superfluid density tensor is then obtained by $n_{sij}=(4m/\hbar^{2}\mathcal{S})[\partial^{2}\Omega(\mathbf{q})/\partial q_{i}\partial q_{j}]_{\mathbf{q}=0}$
\cite{Zhou2012}, where $i,j=(x,y)$ and $\Omega(\mathbf{q})$ is
the thermodynamic potential in the presence of the phase twist. The
critical BKT temperature for the 2D superfluid phase is determined
by (see Supplementary Materials for explanation) \cite{He2012},
\begin{equation}
k_{B}T_{BKT}=\frac{\pi}{2}\mathcal{J}\left(\Delta,Q,\mu,T_{BKT}\right),
\end{equation}
where $\mathcal{J}\equiv[\hbar^{2}/(4m)]\sqrt{n_{sxx}n_{syy}}$ is
the superfluid phase stiffness that does not vanish at $T_{BKT}$.
We note that the above mean-field treatment is applicable in the weakly
interacting regime. The leading higher-order correction to the thermodynamic
potential from pair fluctuations is a spin-wave contribution $\Omega_{\textrm{fl}}=k_{B}T\sum_{\boldsymbol{\mathbf{q}}}\ln[1-e^{-E_{sw}(\mathbf{q})/k_{B}T}]$
with a linear spectrum $E_{sw}(\mathbf{q})$, which is exponentially
small at low temperatures \cite{Yin2014} and therefore is included
in our treatment. Hereafter we set the Fermi wavevector $k_{F}=\sqrt{4\pi N/\mathcal{S}}$
and Fermi energy $E_{F}=\hbar^{2}k_{F}^{2}/(2m)$ as the units for
wavevector and energy, respectively. In all self-consistent calculations,
the interaction parameter is given by $E_{b}=0.2E_{F}$ and the SOC
strength is determined by $\lambda=\sqrt{\lambda_{x}^{2}+\lambda_{y}^{2}}=E_{F}/k_{F}$.
We use $T=0.05T_{F}$ unless otherwise specified. 

Fig. 1 presents the typical low-temperature phase diagram for a Rashba
spin-orbit coupled 2D Fermi gas on the $\Omega_{R}-\delta$ plane.
The case of zero in-plane Zeeman field ($\delta=0$) has been well
explored in the literature \cite{Zhang2008,Sato2009,Sau2010,Oreg2010}.
A topological phase transition is driven by the out-of-plane field
$\Omega_{R}$. The increase of $\Omega_{R}$ will not only change
the topology of the dispersion relation of two helicity branches via
breaking time-reversal symmetry and opening a spin-orbit gap, but
it also induce an effective \textit{p}-wave fermionic pairing in the
lower helicity branch \cite{Zhang2008}. As a result, a gapped topological
superfluid emerges continuously above the threshold $\Omega_{R,c}=\sqrt{\mu^{2}+\Delta^{2}}$.
Associated with this topological phase transition, the energy gap
of the system will first close exactly at $\Omega_{R,c}$ and then
immediately re-open. The presence of a nonzero but small in-plane
field will not change this picture, but it facilitates the finite-momentum
FF pairing due to the Fermi surface deformation in the lower helicity
branch \cite{Liu2013a}. Consequently, a gapped topological FF superfluid
appears, as discussed in the earlier work \cite{Qu2013,Zhang2013}. 

\begin{figure}[t]
\begin{centering}
\includegraphics[clip,width=0.4\textwidth]{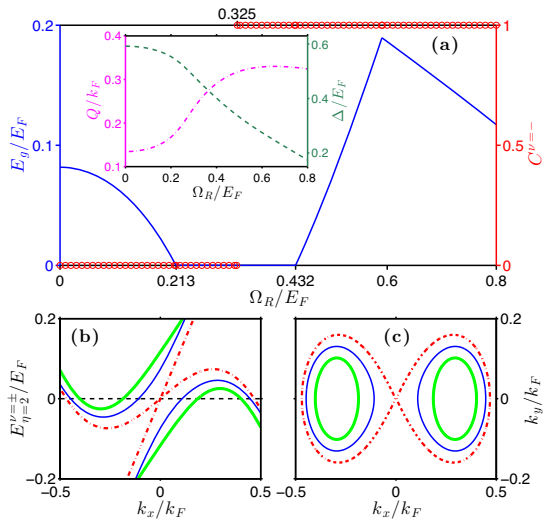} 
\par\end{centering}

\caption{(Color online) (a) Evolution of the minimum excitation gap $E_{g}$
and of the Chern number with increasing $\Omega_{R}$ at a fixed $\delta=0.4E_{F}$.
The inset shows the evolution of the FF pairing momentum $Q$ and
pairing gap $\Delta$. (b) Dispersion relation of the lower helicity
branch $E_{\eta=2}^{\nu}(k_{x},k_{y}=0)$ across the topological transition
at $\Omega_{R}/E_{F}=0.24$ (green thick line), $0.325$ (red dot-dashed
line) and $0.38$ (blue thin line). (c) The corresponding contours
of zero-energy energy spectrum (nodal points) on the $k_{x}$-$k_{y}$
plane.}

\label{fig2} 
\end{figure}

It is appealing to perceive, however, that a \emph{gapless} topological
FF superfluid can also emerge at the sufficiently large in-plane Zeeman
field $\delta\gtrsim0.47E_{F}$. In this case, with increasing $\Omega_{R}$
the Fermi gas is first driven into a gapless state before finally
turns into a gapped topological superfluid, as can be clearly seen
in Fig. \ref{fig2}(a), where we plot the evolution of the minimum
excitation gap \cite{note2}. In the gapless state, the energy of
the lower helicity particle branch ($E_{\eta=2}^{\nu=+}$) becomes
less than zero in a small area slightly away from the origin $\mathbf{k}=0$,
as shown in Fig. \ref{fig2}(b). The nodal points with $E_{\eta=2}^{\nu}(k_{x},k_{y})=0$
form two disjoint loops in the momentum space, see for example Fig.
\ref{fig2}(c), except at a critical value $\Omega_{R}$, where the
two loops connect at $\mathbf{k}=0$. At this value, the topology
of the Fermi surface changes, implying the emergence of a gapless
topological FF superfluid.

\begin{figure}[t]
\begin{centering}
\includegraphics[clip,width=0.4\textwidth]{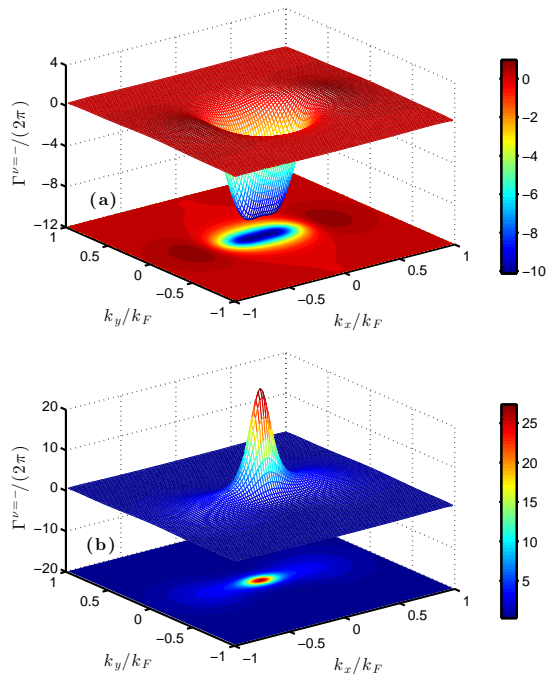} 
\par\end{centering}

\caption{(Color online) Contour plots of the Berry curvature $\Gamma^{\nu=-}\equiv\gamma_{1}^{\nu=-}+\gamma_{2}^{\nu=-}$
before (a, $\Omega_{R}=0.24E_{F}$) and after (b, $\Omega_{R}=0.38E_{F}$)
the topological phase transition at an in-plane field $\delta=0.4E_{F}$.}

\label{fig3}
\end{figure}

To better characterize the topological phase transition in the gapless
state, we calculate the Chern number 
\begin{equation}
C^{\nu=-}=\frac{1}{2\pi}\int d^{2}{\bf k}\sum_{\text{\ensuremath{\eta}}=1,2}\gamma_{\eta}^{\nu=-}
\end{equation}
associated with the hole branches ($\nu=-$), where $\gamma_{\eta}^{\nu}=i(\langle\partial_{k_{x}}\Phi_{\eta}^{\nu}|\partial_{k_{y}}\Phi_{\eta}^{\nu}\rangle-k_{x}\leftrightarrow k_{y})$
is the Berry curvature of the $(\nu,\eta)$ branch \cite{Xiao2010}.
The results are shown in Fig. \ref{fig2}(a) by circles. The topological
state is characterized by a nonzero integer Chern number $C^{\nu=-}=1$.
As anticipated, there is a jump in the Chern number, see for example
the left inset of Fig. \ref{fig2}(a), occurring precisely at the
critical value $\Omega_{R}$ where the topology of the Fermi surface
changes. To gain a concrete understanding of this jump, in Fig. \ref{fig3}
we report the Berry curvature $\Gamma^{\nu=-}\equiv\gamma_{1}^{\nu=-}+\gamma_{2}^{\nu=-}$
just before and after the topological phase transition. In either
case, a sharp peak develops in the $k_{x}$-$k_{y}$ plane around
the origin $\mathbf{k}=0$, pointing downwards or upwards and sitting
on a positive background. In the topologically trivial state, when
atoms scatter on the Fermi surface, a Berry phase $\theta=\int d^{2}\mathbf{k}\Gamma^{\nu=-}\simeq-\pi$
is picked up from the downwards peak but cancels with these accumulated
from the background, leading to a vanishing Chern number. In contrast,
in the topologically non-trivial state, the two contributions are
additive and hence yield a nonzero Chern number $C^{\nu=-}=1$.

\begin{figure}[t]
\begin{centering}
\includegraphics[clip,width=0.48\textwidth]{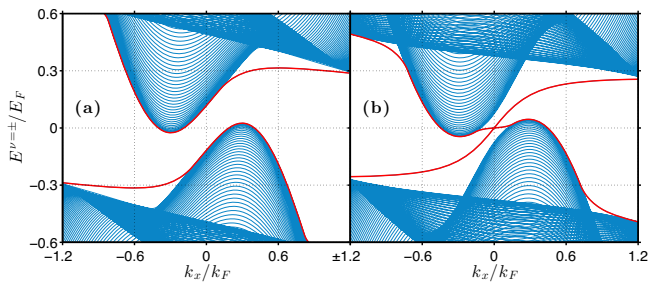} 
\par\end{centering}

\caption{(Color online) Energy spectrum of a 2D strip formed by imposing open
boundary condition (i.e., hard wall confinement) in the \textit{y}-direction.
The edge states due to the confinement are highlighted by the red
lines. The parameters are $\delta=0.4E_{F}$ and $\Omega_{R}=0.24E_{F}$
(a) or $\Omega_{R}=0.38E_{F}$ (b), the same as in Fig. \ref{fig3}.}

\label{fig4} 
\end{figure}

Our gapless topological FF superfluid can support exotic chiral edge
modes, like any topological states \cite{Hasan2010,Qi2011}. To confirm
this fundamental feature, in Fig. \ref{fig4} we report the energy
spectrum of a 2D strip with hard-wall confinement in the \textit{y}-direction.
Two sets of chiral edge states appear due to the confinement, well
localized at the two boundaries, respectively. In the gapless topological
FF state, they come to cross with each other at $k_{x}=0$, giving
rise to two zero-energy Majorana fermions. It is interesting that
the left and right chiral edge states are moving along the \emph{same}
direction at the two boundaries since both states have $v(k_{x})\equiv\partial E_{k_{x}}/\partial k_{x}>0$. 

For conventional gapped topological superfluids, it is known that
the chiral edge states are protected by the non-trivial topology of
the system. For our gapless topological FF superfluids, the chiral
edge states enjoy the same protection, as long as the topological
order of the system is not destroyed by perturbing potentials. To
confirm this, we test the fate of the chiral edge states against a
random disordered potential along the $y$-direction. As discussed
in detail in Supplementary Materials, the low-energy chiral edge states
are found to be very robust. This robustness is a direct consequence
of the topological nature of the system which ensures that turning
on a reasonably strong perturbation cannot immediately destroy the
chiral edge states.

\begin{figure}
\begin{centering}
\includegraphics[clip,width=0.4\textwidth]{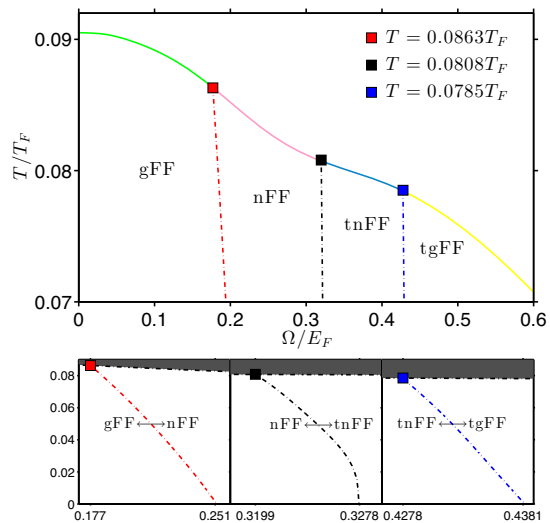} 
\par\end{centering}

\caption{(Color online) Finite-temperature phase diagram as a function of $\Omega_{R}$
at $\delta=0.4E_{F}$. The solid line shows the BKT critical temperature
and the dashed lines give the boundaries between different phases
(see the lower panel for the dependence over the entire temperature
scale).}

\label{fig5} 
\end{figure}

We now turn to discuss the existence of the gapless topological FF
superfluid at finite temperatures, as reported in Fig. \ref{fig5}.
Remarkably, the superfluid density tensor of such superfluids is always
positively defined and is comparable to the total density in magnitude.
This is in sharp contrast to the case of a conventional FF superfluid
in an imbalanced Fermi gas without spin-orbit coupling, whose superfluid
density is precisely zero due to the rotational invariance of the
Fermi surfaces \cite{Yin2014}. As a result, their critical BKT temperature
is very significant. At the typical interaction strength $E_{b}=0.2E_{F}$,
we find that $T_{BKT,tnFF}\simeq0.08T_{F}$, suggesting that it is
indeed within the reach of current experiments.

We also consider a general SOC characterized by $\lambda_{x}=\lambda\cos\psi$
and $\lambda_{y}=\lambda\sin\psi$. We find that the gapless topological
FF superfluid survives over a broad range of the azimuthal angle $\psi$
(see Supplementary Materials). This observation is encouraging, as
the synthetic SOC - to be experimentally realized in cold-atom laboratories
- may not acquire the perfect form of a Rashba SOC. The insensitive
dependence of the gapless topological FF superfluid on a particular
form of SOC therefore means that this conceptually new topological
state of matter is amenable to synthesize with cold atoms.

In summary, we have predicted a gapless topological superfluid with
inhomogeneous Fulde-Ferrell pairing in a two-dimensional spin-orbit
coupled Fermi gas, which possesses gapless excitations in the bulk
as well as non-Abelian Majorana fermions localized at the boundaries.
It exists over a wide range of parameters at finite temperatures and
does not require specific form of spin-orbit coupling, and is therefore
feasible to observe experimentally in cold-atom laboratories. The
gapless excitation in the bulk would lead to richer thermodynamic
and dynamic properties of the system. In three dimensions, the proposed
gapless topological phase may get larger parameter space, as the gapless
feature is favored by high dimensionality. Our work may shed new insights
for the exploration of topological quantum matters, in both cold-atom
\cite{Cheuk2012,Wang2012} and solid-state systems \cite{Alicea2010}. 
\begin{acknowledgments}
This research is supported by the ARC Discovery Projects (FT130100815,
DP140103231 and DP140100637), NFRP-China (2011CB921502 and 2011CB921602)
and NSFC-China (11025421, 11228410, 11175094 and 91221205).

\textit{Note added}. --- In completing this work, we become aware
of a related work \cite{Xu2014}, where the gapless topological Fulde-Ferrell
superfluid in three dimensions is discussed. \end{acknowledgments}


\begin{thebibliography}{10}
\bibitem{Casalbuoni2004}R. Casalbuoni and G. Nardulli, Rev. Mod.
Phys. \textbf{76}, 263 (2004).

\bibitem{Uji2006}S. Uji, T. Terashima, M. Nishimura, Y. Takahide,
T. Konoike, K. Enomoto, H. Cui, H. Kobayashi, A. Kobayashi, H. Tanaka,
M. Tokumoto, E. S. Choi, T. Tokumoto, D. Graf, and J. S. Brooks, Phys.
Rev. Lett. \textbf{97}, 157001 (2006).

\bibitem{Kenzelmann2008}M. Kenzelmann, Th. Strässle, C. Niedermayer,
M. Sigrist, B. Padmanabhan, M. Zolliker, A. D. Bianchi, R. Movshovich,
E. D. Bauer, J. L. Sarrao, and J. D. Thompson, Science \textbf{321},
1652 (2008).

\bibitem{Liao2010}Y.-A. Liao, A. S. C. Rittner, T. Paprotta, W. Li,
G. B. Partridge, R. G. Hulet, S. K. Baur, and E. J. Mueller, Nature
(London) \textbf{467}, 567 (2010).

\bibitem{Gerber2014}S. Gerber, M. Bartkowiak, J. L. Gavilano, E.
Ressouche, N. Egetenmeyer, C. Niedermayer, A. D. Bianchi, R. Movshovich,
E. D. Bauer, J. D. Thompson, and M. Kenzelmann, Nature Phys. \textbf{10},
126 (2014).

\bibitem{Fulde1964}P. Fulde and R. A. Ferrell, Phys. Rev. \textbf{135},
A550 (1964).

\bibitem{Larkin1964}A. I. Larkin and Y. N. Ovchinnikov, Zh. Eksp.
Teor. Fiz. \textbf{47}, 1136 (1994) {[}Sov. Phys. JETP \textbf{20},
762 (1965){]}.

\bibitem{Zwierlein2006}M. W. Zwierlein, A. Schirotzek, C. H. Schunck,
and W. Ketterle, Science \textbf{311}, 492 (2006).

\bibitem{Partridge2006}G. B. Partridge, W. Li, R. I. Kamar, Y.-A.
Liao, and R. G. Hulet, Science \textbf{311}, 503 (2006).

\bibitem{Sheehy2006}D. E. Sheehy and L. Radzihovsky, Phys. Rev. Lett.
\textbf{96}, 060401 (2006).

\bibitem{Hu2006}H. Hu and X.-J. Liu, Phys. Rev. A \textbf{73}, 051603(R)
(2006).

\bibitem{Gubbels2013}K. B. Gubbels and H. T. C. Stoof, Phys. Rep.
\textbf{512}, 255 (2013).

\bibitem{Orso2007}G. Orso, Phys. Rev. Lett. \textbf{98}, 070402 (2007).

\bibitem{Hu2007}H. Hu, X.-J. Liu, and P. D. Drummond, Phys. Rev.
Lett. \textbf{98}, 070403 (2007).

\bibitem{Lin2011}Y.-J. Lin, K. Jiménez-García, and I. B. Spielman,
Nature (London) \textbf{471}, 83 (2011).

\bibitem{Wang2012} P. Wang, Z.-Q. Yu, Z. Fu, J. Miao, L. Huang, S.
Chai, H. Zhai, and J. Zhang, Phys. Rev. Lett. \textbf{109}, 095301
(2012).

\bibitem{Cheuk2012} L. W. Cheuk, A. T. Sommer, Z. Hadzibabic, T.
Yefsah, W. S. Bakr, and M. W. Zwierlein, Phys. Rev. Lett. \textbf{109},
095302 (2012).

\bibitem{Zheng2013}Z. Zheng, M. Gong, X. Zou, C. Zhang, and G.-C.
Guo, Phys. Rev. A \textbf{87}, 031602(R) (2013).

\bibitem{Wu2013}F. Wu, G.-C. Guo, W. Zhang, and W. Yi, Phys. Rev.
Lett. \textbf{110}, 110401 (2013).

\bibitem{Liu2013a}X.-J. Liu and H Hu, Phys. Rev. A \textbf{87}, 051608(R)
(2013).

\bibitem{note1}In the context of superconductors in solid state,
such a FF superfluid is often referred to as a helical state.

\bibitem{Barzykin2002}V. Barzykin and L. P. Gor'kov, Phys. Rev. Lett.
\textbf{89}, 227002 (2002).

\bibitem{Dong2013}L. Dong, L. Jiang, H. Hu, and H. Pu, Phys. Rev.
A \textbf{87}, 043616 (2013).

\bibitem{Shenoy2013}V. B. Shenoy, Phys. Rev. A \textbf{88}, 033609
(2013).

\bibitem{Hasan2010}M. Z. Hasan and C. L. Kane, Rev. Mod. Phys. \textbf{82},
3045 (2010).

\bibitem{Qi2011}X.-L. Qi and S.-C. Zhang, Rev. Mod. Phys. \textbf{83},
1057 (2011).

\bibitem{Nayak2008}C. Nayak, S. H. Simon, A. Stern, M. Freedman,
and S. Das Sarma, Rev. Mod. Phys. \textbf{80}, 1083 (2008).

\bibitem{Mourik2012}V. Mourik, K. Zuo, S. M. Frolov, S. R. Plissard,
E. P. A. M. Bakkers, and L. P. Kouwenhoven, Science \textbf{336},
1003 (2012).

\bibitem{Williams2012}J. R. Williams, A. J. Bestwick, P. Gallagher,
S. S. Hong, Y. Cui, Andrew S. Bleich, J. G. Analytis, I. R. Fisher,
and D. Goldhaber-Gordon, Phys. Rev. Lett. \textbf{109}, 056803 (2012).

\bibitem{Rokhinson2012}L. P. Rokhinson, X. Liu and J. K. Furdyna,
Nature Phys. \textbf{8}, 795 (2012).

\bibitem{Zhang2008}C. Zhang, S. Tewari, R. M. Lutchyn, and S. Das
Sarma, Phys. Rev. Lett. \textbf{101}, 160401 (2008).

\bibitem{Sato2009}M. Sato, Y. Takahashi, and S. Fujimoto, Phys. Rev.
Lett. \textbf{103}, 020401 (2009).

\bibitem{Sau2010}J. D. Sau, R. M. Lutchyn, S. Tewari, and S. Das
Sarma, Phys. Rev. Lett. \textbf{104}, 040502 (2010).

\bibitem{Oreg2010}Y. Oreg, G. Refael, and F. von Oppen, Phys. Rev.
Lett. \textbf{105}, 177002 (2010).

\bibitem{Liu2012a}X.-J. Liu, L. Jiang, H. Pu, and H. Hu, Phys. Rev.
A \textbf{85}, 021603(R) (2012).

\bibitem{Liu2012b}X.-J. Liu and H. Hu, Phys. Rev. A \textbf{85},
033622 (2012).

\bibitem{Wei2012}R. Wei and E. J. Mueller, Phys. Rev. A \textbf{86},
063604 (2012).

\bibitem{Ruhman2014}J. Ruhman and E. Altman, arXiv:1401.7343.

\bibitem{Chen2013}C. Chen, Phys. Rev. Lett. \textbf{111}, 235302
(2013).

\bibitem{Liu2013b}X.-J. Liu and H. Hu, Phys. Rev. A \textbf{88},
023622 (2013).

\bibitem{Qu2013}C. Qu, Z. Zheng, M. Gong, Y. Xu, L. Mao, X. Zou,
G. Guo, and C. Zhang, Nat. Comm. \textbf{4}, 2710 (2013).

\bibitem{Zhang2013}W. Zhang and W. Yi, Nat. Comm. \textbf{4}, 2711
(2013).

\bibitem{Alicea2010}J. Alicea, Phys. Rev. B \textbf{81}, 125318 (2010).

\bibitem{Hu2011}H. Hu, L. Jiang, X.-J. Liu, and H. Pu, Phys. Rev.
Lett. \textbf{107}, 195304 (2011).

\bibitem{Zhou2012}K. Zhou and Z. Zhang, Phys. Rev. Lett. \textbf{108},
025301 (2012).

\bibitem{He2012}L. He and X.-G. Huang, Phys. Rev. Lett. \textbf{108},
145302 (2012).

\bibitem{Yin2014}S. Yin, J.-P. Martikainen, and P. Törmä, Phys. Rev.
B \textbf{89}, 014507 (2014).

\bibitem{note2}Quite generally, a two-component Fermi superfluid
will become gapless in the limit of a large in-plane Zeeman field.
For a detailed discussion, we refer to Supplemental Materials.

\bibitem{Xiao2010}D. Xiao, M.-C. Chang, and Q. Niu, Rev. Mod. Phys.
\textbf{82}, 1959 (2010).

\bibitem{Xu2014}Y. Xu, R. Chu, and C. Zhang, Phys. Rev. Lett. \textbf{112},
136402 (2014).\end{thebibliography}
\end{document}